 \documentclass[letterpaper,12pt,nofootinbib,aps,tightenlines]{revtex4}

\usepackage{amsmath,amsfonts,amssymb}
\usepackage{mathrsfs}
\usepackage{graphicx}
\usepackage[english]{babel}

\usepackage{wrapfig}
\usepackage[rightcaption]{sidecap}

\usepackage[left]{lineno}
\usepackage{mathtools}

\newcommand{\be}{\begin{equation}}
\newcommand{\ee}{\end{equation}}
\newcommand{\bes}{\begin{equation*}}
\newcommand{\ees}{\end{equation*}}
  
\usepackage{color}

\begin{document}
\setcounter{page}{0}
 
\title{Astro2020 Science White Paper: \\
Cosmology with a Space-Based Gravitational Wave Observatory}

\vspace{1.0cm}

\begin{abstract}

There are two big questions cosmologists would like to answer -- How does the Universe work, and what are its origin and destiny? A long wavelength gravitational wave detector -- with million km interferometer arms, achievable only from space -- gives a unique opportunity to address both of these questions. A sensitive, mHz frequency observatory could use the inspiral and merger of massive black hole binaries as standard sirens, extending our ability to characterize the expansion history of the Universe from the onset of dark energy-domination out to a redshift $z \sim 10$. A low-frequency detector, furthermore, offers the best chance for discovery of exotic gravitational wave sources, including a primordial stochastic background, that could reveal clues to the origin of our Universe.

\end{abstract}

\date{\today}

\maketitle

\noindent
Thematic Science Area: Cosmology and Fundamental Physics

\vspace{0.3cm}
\noindent
Principal Author: Robert Caldwell\\
Institution: Dartmouth College\\
Email: robert.r.caldwell@dartmouth.edu\\
Telephone: +1 (603) 646-2742

\vspace{0.3cm}
\noindent
Co-Authors: \\
Mustafa Amin (Rice University),\\
Craig Hogan (University of Chicago), \\
Kelly Holley-Bockelmann (Vanderbilt University),\\
Daniel Holz (University of Chicago), \\
Philippe Jetzer (University of Zurich, Switzerland), \\
Ely Kovetz (Johns Hopkins University), \\
Priya Natarajan (Yale University),\\
David Shoemaker (Massachusetts Institute of Technology),\\
Tristan Smith (Swarthmore College), \\
Nicola Tamanini (Max Planck Institute for Gravitational Physics, Germany)

\newpage

\noindent{\bf Big Questions}
 
The ground-breaking detection of gravitational waves (GWs) by the LIGO Scientific and Virgo collaborations \cite{Abbott:2016blz} marks the beginning of the era of GW astronomy. With just a handful of events, GW astronomers have begun to peer into the dark Universe. The exciting, first results already include new insights into black hole and neutron star physics, new tests of general relativity, and new measures of the cosmos. Yet, even as LIGO and Virgo are implementing upgrades to extend their reach, forward-thinking astronomers are looking ahead towards new ways to probe the Universe.

The low frequency, mHz GW spectrum is a new frontier for cosmology. The brightest mHz sources are the mergers of massive black hole binaries (MBHB), which span redshifts $z\sim 1-10$. These events can be used as standard sirens to chart the cosmic expansion history all the way back to the onset of dark energy and before. The mHz frequency range may also contain clues to the origin of the matter-antimatter asymmetry of the Universe. GWs released in an electroweak phase transition, which generates the asymmetry when the temperature of the cosmic fluid was around $1-100$~TeV, would redshift to mHz by today. Furthermore, sensitivity to a stochastic GW background improves at low frequency, assuming fixed strain sensitivity. This means if there is a relic primordial background of GWs from the very early Universe, the best chance for detection is at low frequencies.  Clearly, a GW observatory with peak sensitivity in the mHz range has the potential to help address profound questions in cosmology including the nature of dark energy and the origin of matter. Such an observatory offers unprecedented opportunity for discovery of new physics stretching all the way back to the origin of the Hot Big Bang.

The Laser Interferometer Space Antenna (LISA) is a proposed space-borne GW detector that will open a new vista on gravitational astrophysics and cosmology in the mHz frequency band.  LISA was selected by the European Space Agency (ESA) in June 2017 for ESA's third large (L3) mission in the ``Cosmic Vision" plan \cite{ESALISA}. As proposed, LISA \cite{Audley:2017drz} will consist of a trio of spacecraft separated by millions of kilometers in an Earth-trailing orbit. These spacecraft will detect GWs in space by using precision laser interferometry to track relative changes in spacecraft separation. Important technology milestones have already been surpassed by the LISA Pathfinder mission \cite{Armano:2016bkm,Armano:2018kix}, demonstrating the viability of the LISA science goals. The nominal mission lifetime is four years, with an expected launch date in the early 2030s. The primary science objectives include measurement of the cosmic expansion history and search for stochastic backgrounds from TeV scale physics and the very early Universe. Such an instrument represents a {\it tremendous opportunity for cosmology.}

\vspace{\baselineskip}
\noindent{\bf Cosmography}

Measurement of distance underlies two profound scientific discoveries, the cosmic expansion and acceleration. GWs now offer a unique, purely gravitational way to measure distance. The new method uses absolutely calibrated distances determined by measuring the waves generated by binary inspirals and mergers. These binaries are completely characterized by a well-understood mathematical model, and their gravitational waveforms as they spiral together are known from first principles. These {\it standard sirens} have recently been used for the first gravitational measurement of the Hubble constant, using the famous LIGO/Virgo binary neutron star event, GW170817 \cite{Abbott:2017xzu}. Combination with follow-up electromagnetic (EM) observations \cite{Hotokezaka:2018dfi} has refined the measurement of $H_0$. In the coming decade, we can expect ground-based detectors to build a gravitational Hubble diagram out to $z \sim 0.5$ \cite{Chen:2017rfc} that will break the tension between local (distance ladder) and distant cosmic microwave background (CMB) measurements of $H_0$ \cite{Freedman:2017yms}.  A long-wavelength, space-based observatory has the capability to extend this reach to distances which have never been probed and address questions that are impossible with purely EM observations.

\begin{SCfigure}[0.99][t]
\caption{The redshift range of the expansion rate probed by various cosmological methods are illustrated. The loudest objects in the sky seen by a mHz GW observatory will be massive black hole binaries, ranging from $z \simeq 1 - 10$  (red band) with a mean redshift $\bar z\simeq 3$ (large dot) \cite{Tamanini:2016uin}. Representative mean redshifts for other techniques  \cite{Spergel:2013,Aghamousa:2016zmz,Zhan:2017uwu,Aasi:2013wya} -- type Ia SNe, LSS (Baryon Acoustic Oscillations and Weak Lensing), and BNS (ground-based standard sirens) -- are shown by black dots. The line $q=0$ marks the onset of acceleration. }
\includegraphics[width=0.56\textwidth]{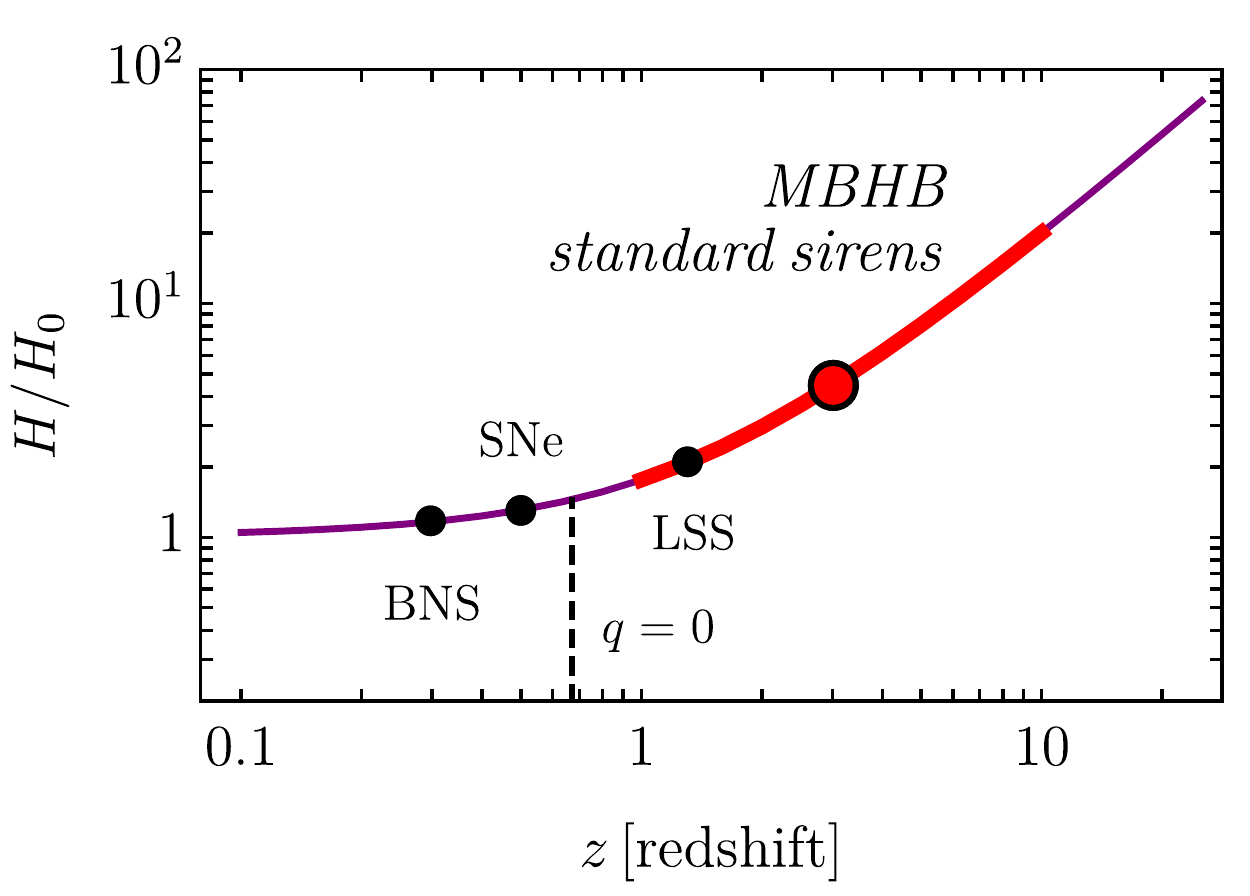}
\label{fig:SirensFig}
\end{SCfigure}

The basic method of estimating distances from measured waveforms is elegantly simple \cite{Schutz:1986gp}: the chirp time $\tau$ of an inspiral or merger waveform, together with its orbital frequency $\omega$ and strain $h$, gives an absolute luminosity distance $D\approx c/(\omega^2\tau h)$, with a numerical factor depending on details of the configuration that are precisely determined by the properties of the measured waveform. Roughly speaking, the directly-measured wave period gives the redshifted final Schwarzschild radius, so the ratio of that length to the directly-measured metric strain, $h$, gives the luminosity distance. To measure the expansion history, the redshift is then required.

Among the rich astrophysical sources expected to produce a detectable GW signal at mHz frequencies, only MBHBs with masses ranging from $10^4$ to $10^7$ solar masses will produce a detectable EM counterpart from which the redshift can be obtained. These systems are expected to merge in a gas rich environment and are abundant at very high redshifts, making them ideal standard sirens to probe the cosmic expansion at distances other techniques cannot reach. 

Measurement of these inspiral waves makes it possible to directly determine the luminosity distance to a single source with an intrinsic precision that in favorable cases can be as good as $0.1\%$ \cite{Holz:2005df}. This is the ``raw" value achievable given only instrumental limitations; in reality, effects such as weak gravitational lensing will contribute an important systematic at high redshift. In the coming decade, however, we expect to have high-precision maps of the intervening large-scale structure from the Large Synoptic Survey Telescope (LSST) \cite{LSSTurl} and the Wide-Field Infrared Survey Telescope (WFIRST) \cite{WFIRSTurl} which can be used to ``de-lens" or compensate for magnification bias \cite{Hirata:2010ba}. We will also rely on the large telescopes online in the era of LISA for EM follow-up. Detailed forecasts show that a space-based GW observatory will return valuable information about the expansion history and laws of gravitation on cosmic scales that are inaccessible by other means \cite{Tamanini:2016zlh,Caprini:2016qxs,Belgacem:2017ihm,Ezquiaga:2018btd}. As a timely example, quasar luminosity distances have recently been claimed to show a deviation from the $\Lambda$CDM expansion history at $z\sim3$ \cite{Risaliti:2018reu}. A quick analysis shows that LISA will be able to distinguish the quasar luminosity distance claim from $\Lambda$CDM at high significance \cite{Caldwell:2019}.

We caution that LISA is not a dark energy machine: other experiments will measure the low-redshift dark-energy parameters with a greater figure of merit, as defined by the Dark Energy Task Force \cite{Albrecht:2006um}, than is achievable using MBHB standard sirens. The strength of LISA cosmography is the range of redshift and the purely gravitational method. After the Dark Energy Spectroscopic Instrument (DESI) \cite{DESIurl}, LSST, and WFIRST there will be good measurements of the expansion history at $z\lesssim 3$ but not beyond. (Intensity mapping may reach higher redshift, but is in its infancy \cite{Ansari:2018ury}.) There is very good theoretical reason to explore the epoch $z\gtrsim 3$: theoretical explanations for the standard $\Lambda$CDM cosmology are typically fine-tuned, but quintessence and other alternatives that alleviate the fine-tuning issues typically predict signatures at $z\gtrsim 2$.  Measurement of the expansion history at these redshifts will put dark energy theories to the test. 

By the time that such an observatory will be ready to launch, there will be a new generation of ground-based GW detectors online. (See Ref.~\cite{Barsotti} for an overview.) Coordination among these facilities will tighten the standard siren Hubble diagram across a range of redshift. For example, up to hundreds of stellar mass black hole binaries will be resolvable in the mHz range, allowing weeks advance notice of GW and EM probes of the merger, forecasted to seconds accuracy and localized within a square degree on the sky \cite{Sesana:2016ljz}. Similarly, data from the ground will enhance the sensitivity of LISA and boost the detection rate \cite{Wong:2018uwb}, thereby enabling multi-frequency studies of GWs.

(See also ``Astro2020 Science White Paper: What we can Learn from Multi-band Gravitational-Wave Observations of Black Hole Binaries" \cite{Cutler:2018}.)

The long path length to high redshift GW standard sirens can be used to probe large-scale structure, too. The detection of strongly-lensed GW sources provides an additional tool for cosmography \cite{Seto:2003iw,Sereno:2010dr,Sereno:2011ty,Kyutoku:2016zxn}. Multiple images (more precisely the chirping GWs) and time-delay measurements, coupled with high calibration accuracy and annual motion of the space-based observatory will enable the localization of binaries. Lensing amplification will help further to find the host galaxies. Since most of the optical depth for lensing is provided by intervening massive galactic halos, the detection of multiple events will provide information on evolution and formation of large-scale structure, constraints on cosmological parameters, and tests of competing theories of gravity. 

(See also ``Astro2020 Science White Paper: Tests of General Relativity and Fundamental Physics with Space-based Gravitational Wave Detectors" \cite{Berti:2018}.)

\vspace{\baselineskip}
\noindent{\bf Discovery}
  
GWs in the Universe today preserve a record of macroscopic mass-energy flows over vast stretches of cosmic history, and can be used to probe aspects of new physics never before explored.  A sensitive low-frequency detector opens up an enormous range of discovery space. Here we review some ideas at the frontiers of physics that give a flavor of the type of discoveries one might hope to make.

\begin{figure}[b]
    \includegraphics[width=0.95\linewidth]{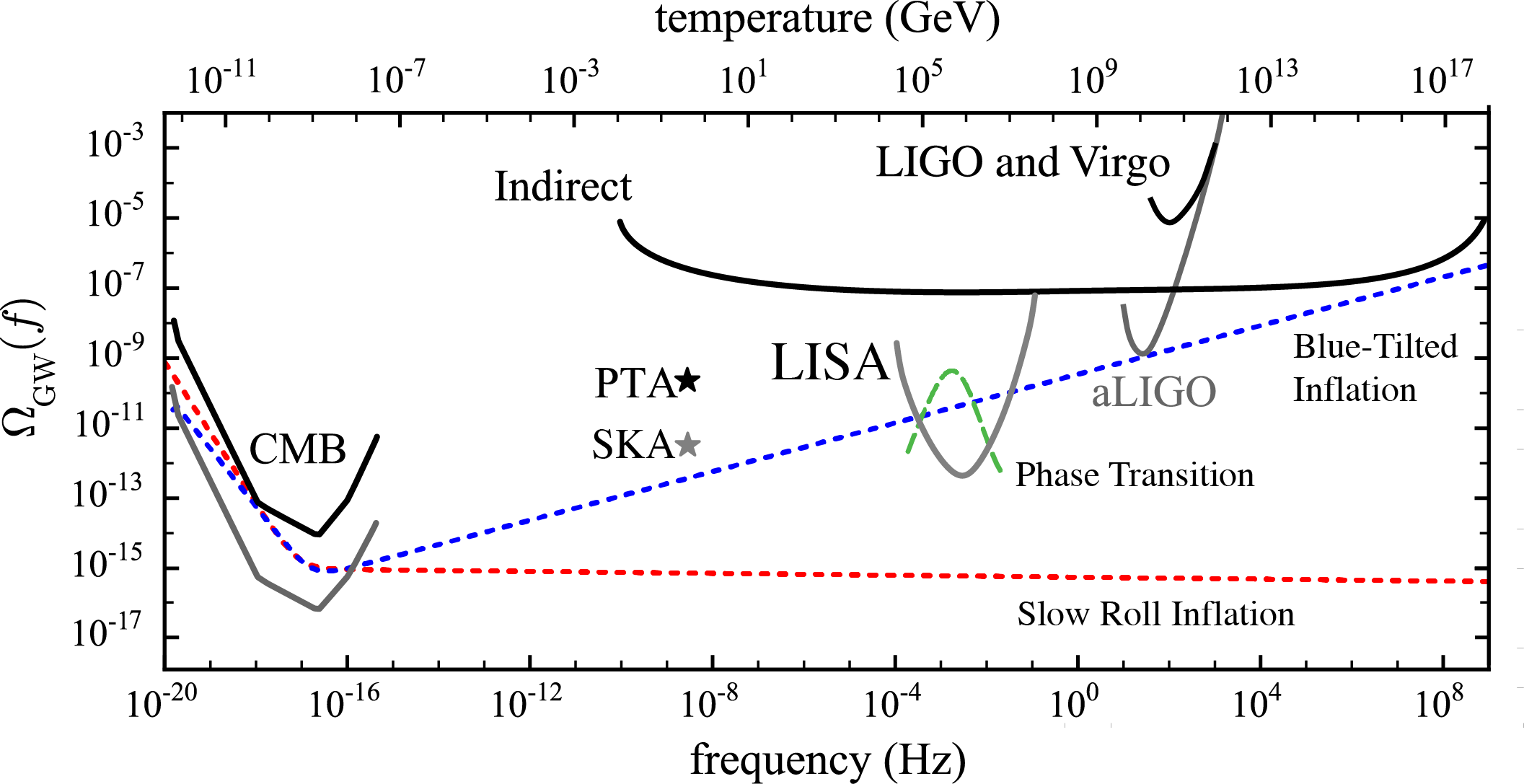}    
\caption{Gravitational Wave Cosmology: The sensitivity of LISA to a stochastic background is shown, relative to other observatories and methods. The dashed red and blue curves show the prediction of a standard inflationary model and the maximum blue-tilted spectrum. The long-dashed green curve is an example of the spectrum from a strongly first-order electroweak phase transition (case 1 from Ref.~\cite{Caprini:2015zlo}). Black and gray curves show current and future sensitivity. The LIGO and Virgo curves show the sensitivity level at the time of the first detections of GWs; aLIGO shows the projected sensitivity of the advanced LIGO design; PTA shows the Pulsar Timing Array sensitivity; SKA shows the projected sensitivity by the Square Kilometer Array \cite{Sesana:2010mx}; CMB shows the bound $r<0.1$ (95\% C.L.) via B mode polarization, whereas the lower, gray curve shows a projected sensitivity $r<0.001$ for future CMB experiments; Indirect is based on Big Bang Nucleosynthesis and CMB sensitivity to additional light degrees of freedom. Figure adapted from Ref.~\cite{Lasky:2015lej}.}
\label{fig:NoiseModel}
\end{figure}

$\bullet$  A stochastic GW background (SGWB) is predicted in a wide range of particle physics-based theories of the early Universe. We illustrate several possibilities in the context of current and future GW observatories in Fig.~\ref{fig:NoiseModel}, relative to the 4-year LISA sensitivity curve for a power-law stochastic background \cite{Thrane:2013oya}. Any power-law SGWB that intersects the LISA curve is detectable at $5\sigma$ or greater \cite{LISAscird,Caldwell:2018giq}.

A burst of inflation, the bang of the Hot Big Bang, may have produced a spectrum of relic GWs. These are quantum fluctuations of the gravitational field that are amplified and stretched by exponential expansion to macroscopic scales. The imprint of these primordial waves on the CMB is under hot pursuit. Given the current bounds on such a background, from the upper limit on CMB B-mode polarization \cite{Ade:2018gkx}, we extrapolate the predictions of slow roll inflation to the mHz band. As shown in the figure, this spectrum is hopelessly far below the sensitivity of LISA. However, intriguing inflationary scenarios, based on new ideas about symmetry breaking, predict blue-tilted spectra \cite{Adshead:2013qp,Dimastrogiovanni:2016fuu,Fujita:2018ehq,Maleknejad:2018nxz}. The maximum blue-tilted spectrum, illustrated in Fig.~\ref{fig:NoiseModel}, will be within range of LISA \cite{LISAscird,Caldwell:2018giq}. Detection of such a signal would be a profound discovery, revealing the origin of the large-scale structure of the Universe and the particle physics that gave rise to the primordial fireball.

A broken spectrum of GWs, peaked near mHz frequencies, is the signature of an electroweak-scale phase transition. In such a scenario, the physical vacuum once had a significantly higher free energy that is liberated in a phase transition to a final, true vacuum and eventually converted into thermal energy of radiation and hot plasma \cite{Witten:1984rs,Hogan:1986qda,Hindmarsh:2017gnf}. Such a phase transition could play a key role in determining the prevalence of matter over antimatter in the Universe. The potential for a space-based GW observatory to detect the SGWB produced by a strongly first-order cosmological phase transition is a subject of intense study \cite{Caprini:2015zlo}. The stochastic background depends on a broad range of physics, including the collisions of bubbles, sound waves, and fluid turbulence. Fig.~\ref{fig:NoiseModel} shows an example of such a scenario wherein new physics at the TeV scale leads to a background of mHz GWs. The top scale in the figure gives the temperature of the cosmic fluid in the early Universe when a GW of a given present-day frequency crossed the Hubble radius. Because the characteristic bubble size is typically at least an order of magnitude smaller than the horizon radius, the translation from temperature to frequency is similarly offset by at least an order of magnitude. That means a phase transition within reach of a mHz GW observatory would also be within reach of the next generation of particle accelerators, providing a unique window to phenomena at energy scales from $10$~GeV up to $100$~TeV  \cite{Benedikt:2018ofy,Tang:2015qga}.

$\bullet$  GWs provide the only way to directly probe the interactions of primordial-black-hole dark matter \cite{Kovetz:2017rvv}. Primordial black holes can dynamically form binaries, typically resulting in highly eccentric orbits at birth~\cite{Ali-Haimoud:2017rtz}. A long wavelength detector can be invaluable in distinguishing this formation channel from a stellar origin through measurements of the spin and eccentricity~\cite{Cholis:2016kqi} as well as the mass spectrum~\cite{Kovetz:2016kpi}. Primordial-black-hole binary inspirals and mergers, as well as hyperbolic encounters also contribute a stochastic background that would be accessible at mHz frequencies \cite{DeVittori:2012da,Clesse:2016ajp,Kuhnel:2017bvu,Guo:2017njn,Garcia-Bellido:2017knh}. If tiny, $10^{20}-10^{22}$~g primordial-black-hole dark matter forms out of large density fluctuations in the early Universe, the concomitant spectrum of GWs will peak in the mHz band and be detectable by LISA \cite{Cai:2018dig,Bartolo:2018evs}.

$\bullet$ If some of our most basic notions are valid about the important role of symmetries underlying a unification of forces, then the Universe may be filled with exotic relics such as cosmic strings \cite{Kibble:1976sj} or superstrings \cite{Polchinski:2004hb}. These are microscopically thin, astronomically long objects that writhe and twist under enormous tension, emitting bursts and backgrounds of gravitational radiation. GWs from individual string loops, which emit in a perfect harmonic series, can be identified by extremely narrow GW ``spectral lines" \cite{DePies:2009mf}. Other distinctive bursts, recognizable by a universal wave-form, can be seen from loops that beam GWs in our direction when a momentary ``cusp" event produces a sharply-bent segment of string moving at nearly the speed of light \cite{Key:2008tt}.

(See also ``Astro2020 Science White Paper: The Discovery Potential of Space-Based Gravitational Wave Astronomy" \cite{discoWP}.)
  
\vspace{\baselineskip}
\noindent{\bf Summary}

Low frequency GWs carry unique cosmological information. We anticipate new cosmographic measurements with standard sirens that reveal the physics of cosmic acceleration. We foresee tremendous potential for the discovery of new, exotic sources. Beneath the foreground of unresolved astrophysical GW sources, we have a chance to detect a stochastic background and peer back to the early Universe.  Of course, there are many more ways to use GW observations for cosmology than covered in this short document.
The mHz GW spectrum is a new frontier.
      
\vfill

\eject



\end{document}